\documentclass{appolb}%
\usepackage{graphicx}
\usepackage{amsmath}

\def\vec#1{\ensuremath{\mathchoice
                     {\mbox{\boldmath$\displaystyle#1$}}
                     {\mbox{\boldmath$\textstyle#1$}}
                     {\mbox{\boldmath$\scriptstyle#1$}}
                     {\mbox{\boldmath$\scriptscriptstyle#1$}}}}

\begin{document}
%
\title{The role of scalar and mass  interactions in 
a relativistic model of the charmonium spectrum 
}
\author{M. De Sanctis \footnote{mdesanctis@unal.edu.co}
\address{Universidad Nacional de Colombia, Bogot\'a, Colombia }
\\
}
\maketitle
\begin{abstract}
Lorentz scalar and mass interactions are studied in more detail in
the framework of a reduced Dirac equation for heavy quark-antiquark mesons.
A microscopic model for these interactions is proposed and analyzed.
The charmonium mass spectrum is reproduced by means of two free parameters;
a third parameter is fixed  by means of
a phenomenological hypothesis in accordance with the model.

\end{abstract}
\section{Introduction}\label{intro}
The study of hadronic spectroscopy still represents a challenge in theoretical physics,
considering that Quantum Chromo-Dynamics (QCD) cannot be applied directly 
to this study and
numerical lattice simulations require huge computational efforts.\\
On the other hand, when developing phenomenological models,   
many different aspects must be  taken into account at the same time.
In the framework of a constituent model with a fixed number of valence  quarks,
one has to face primarily the problem of implementing a relativistic wave equation 
for the hadronic system;
within the relativistic equation model,
it is also necessary to select a suitable interaction for the quarks.\\
We stress once again that a completely consistent relativistic
two-body (or many-body) wave equation does not exist but, on the other hand,
it necessarily represents the starting point for the study of hadronic spectroscopy.\\ 
A series of works has been previously developed by the author with the aim 
of constructing a consistent model for that phenomenology.\\
In particular,  in the work  \cite{localred},   
a reduced Dirac-like equation (RDLE) was introduced
for studying the spectroscopy of quark composed systems.
This equation has a local form in the coordinate space.
Furthermore, our RDLE is particularly suitable when a vector plus scalar
interaction is considered. 
The same procedure of reduction was also applied to other relativistic equations 
obtaining very similar numerical results.
An accurate calculation of the charmonium spectrum was performed using a small
 number of free parameters in Ref. \cite{rednumb}.
Furthermore,
in a subsequent work \cite{relvar}, the Lorentz structure of the interaction terms was
studied in more detail, considering a covariant form of the  relativistic equation
of the model.\\
In  all those works 
a specific form of regularized \textit{vector interaction} was used.
That interaction had been
 introduced and studied previously in Ref. \cite{chromomds}.
We highlight here that a vector interaction alone is not sufficient to give an accurate
 reproduction of the charmonium spectrum.
To this aim, the contribution of a
\textit{scalar interaction}  was included.\\
Starting from this observation,  in  the present study we analyze in more detail
the scalar  interaction.
We also consider the possibility of replacing the scalar interaction 
with a \textit{mass interaction}.\\
Incidentally, we note that the use of a fully relativistic model with
an accurate interaction may also help to study the main properties
of higher excitation states in order to understand if these states 
can be described as quark-antiquark states or further 
(exotic) physical effects must be taken into account \cite{fersan}.\\
\vskip 0.1 truecm
\noindent
Going back to the present model,
we also recall that, while the vector interaction  can be related
to an effective reduction of the one-gluon exchange QCD interaction, 
in the scalar and mass case one should take into account, 
by means of specific techniques, many-gluon exchange processes.\\
Furthermore, also in the framework of a rigorous derivation of a nonrelativistic
potential, 
obtained for example by matching the short-distance perturbative part to long-distance 
lattice QCD results \cite{las}, 
there is no clear way to distinguish between the contributions given by the time
component of a vector interaction and those given 
by a hypothetical effective scalar interaction.\\
In any case, within our model, the scalar and the mass interactions can be considered
phenomenological interactions given by an underlying field that has the same quantum numbers of the vacuum.
This field is necessary not only to reproduce  the charmonium 
spectrum but also (for the consistency of the model)
to compensate the vector repulsive quark self-energy with an attractive counterterm, 
as determined by the \textit{energy balance}, given in Eq. (23) 
of Ref. \cite{rednumb}.
In the present work the balance equation will be generalized 
to include the case of  the mass interaction, as it will be shown
in Eqs. (\ref{genbalance1}) and (\ref{genbalance2}) of Sect. \ref{modint}.
\vskip 0.5 truecm
\noindent
The problem of a scalar quark interaction has been faced in many different ways.
We quote here only a few works that may be partially related to the present study.
Historically, a scalar interaction, quartic with respect to the fermionic field,
was introduced in the Nambu Jona-Lasinio (NJL) model \cite{njl1}, \cite{njl2}.
This model was extensively applied to the study of quark matter interactions,
also highlighting possible unusual bound states in quark matter
\cite{mish1}, \cite{mish2}.\\
The NJL model has been studied and modified
in Ref. \cite{meg1} including a running coupling obtained by a fractal
approach to QCD.\\
A nonlinear Klein-Gordon equation for the study of condensates in hadronic particles
has been proposed in Ref. \cite{meg2}.\\
However, considering the difficulty of applying the NJL approach to our relativistic
equation for a two-body bound state, we shall not use directly this method 
in the present work, preferring a standard relativistic interaction 
with a spatial phenomenological potential.\\
We recall that the role of a scalar interaction, 
in the case of two interacting scalar particles,
was deeply studied by means of a model quantum field theory 
\cite{dare1},\cite{dare2}.\\
The \textit{linear sigma model} was introduced for the interaction of 
hadronic particles considering the exchange of the (scalar) sigma meson.
This model, considered as an effective field theory \cite{appel},
was also applied to the quark scalar interaction.
For a relatively recent application, see for example Ref. \cite{das}.  
In the present work, when  studying the underlying structure of the scalar
and mass interaction, we establish, in a different way, a tentative connection with
the first scalar hadronic resonances, now denoted as $f_0(500)$ (formerly sigma meson )
and $f_0(980)$.

\vskip 0.1 truecm
\noindent
The present  phenomenological study has been developed 
taking into account the complexity of the quark interaction,
with no intent to draw definitive conclusions. 
To avoid repetitions, the reader is frequently  referred to the previous works
\cite{localred,rednumb,relvar,chromomds} that can  also help to gain 
a better understanding of the whole subject.
The contents of the present paper are organized as follows.
In Subsect. \ref{symbnot} the symbols and notations of the work 
are introduced.
In Sect. \ref{lorstruct}, the Lorentz structure of the scalar interaction is revised
and the mass interaction is introduced.
In Sect. \ref{charmspectr} a general discussion about the reproduction of the charmonium spectrum is given.
In Sect. \ref{modint} we construct the scalar and mass interaction of the model, 
starting from an elementary interaction of point-like particles, then introducing
finite density distributions for the interacting quarks.
The corresponding self-energies are also determined and the balance equation is
generalized to the case of the mass interaction.
In Sect. \ref{orint}, we try to construct the scalar and mass interactions
introducing an underlying scalar field in order  
to improve the consistency of the model and to interpret the  physical meaning 
of its  parameters. 
Finally, the results are summarized and discussed.
In the Appendix \ref{reductm}, we give,
for the mass interaction, the reduced expressions to be inserted in the RDLE
 of the model.



\subsection{Symbols and Notation}\label{symbnot}
\noindent
The following notation is used in the paper:
\begin{itemize}

\item  
The invariant product between four vectors is standardly written as: 
$V^\mu U_\mu=V^0 U^0- \vec V \cdot \vec U$.\\

\item
The lower index $i (j)=1,2$ is the \textit{particle index},
referred to the quark and to the antiquark.

\item
The Dirac wave functions will be represented by the letter $\Psi$.\\ 

\item
The subindex $X$ will used to indicate, 
for different quantities,
the scalar ($X=S$) or mass ($X=M$) character of the corresponding interaction.
In the text, we shall also write, in general,
 ``x-interaction", ``x-charge density", etc.\\

\item
Furthermore, the subindex $E$ (always associated to $X$) will be used to indicate,
for a given quantity, 
the `` elementary'' or ``point-like" character of the corresponding x-interaction.\\

\item
For the \textit{general} case of two different x-charge densities, 
in Sect. \ref{modint},  the subindex $G$ will be used.\\

\item
Throughout the work we use the standard natural units, 
that is $\hbar=c=1$.\\

\end{itemize}

\vskip 1.0 truecm
\section{ The Lorentz structure of the scalar and mass interaction}\label{lorstruct}
We analyze here the Lorentz structure of the \textit{scalar} and \textit{mass}
interactions.\\
In the first case we have, for a two-body scalar interaction, 
the following standard expression
\begin{equation}\label{scallor}
{\cal V}_S=
\bar \Psi V_S(r)\Psi= 
\Psi^\dag\gamma^0_1\gamma^0_2 V_S(r)\Psi~. 
\end{equation}
The scalar character of this term is obvious. 
This interaction, 
given   in Eq. (11) of Ref. \cite{rednumb}
was used in that work to study in detail the charmonium spectrum.
For the potential  we used there the notation $V^S_{(2)}(r)$
to indicate its two-body character. Here, for the same potential, 
we simply write $ V_S(r)$.\\
We recall that the distance $r$ between the quark and the antiquark is defined in the  center of mass reference frame (CMRF), that will be always used in this work.\\
In more detail,
as shown in Eq. (30) of Ref. \cite{relvar}, that interaction can be written in the
momentum space, with the  CMRF momentum transfer of Eq. (31); 
then, the covariant integration of Eq. (25) is performed, leading to 
the covariant integral equation shown in Eq. (34) of the same work.

\vskip 0.5 truecm
\noindent
On the other hand,
the \textit{mass interaction} operator can be formally introduced 
by means of the following substitution in the two-body Dirac equation
\begin{equation}\label{massubst}
m_i \rightarrow m_i + U_i^M(r)~.
\end{equation}
For the charmonium, we have $ m_i=m_q$.
Symmetry with respect to interchange of $c$ and $\bar c$ requires
\begin{equation}\label{v_m}
U_i^M(r)={\frac {V_M(r)} {2}}~.
\end{equation}
By using Eqs. (34), (35) and (36) of Ref. \cite{relvar}, one finds that
the corresponding mass interaction term, in the covariant form of the equation, 
takes the form
\begin{equation}\label{masscov}
{\cal V}_M=
 \bar \Psi {\frac {V_M(r)} {2} }
{\frac {P_\mu  (\gamma_1^\mu+\gamma_2^\mu)} {M} } \Psi ~.
\end{equation}
For the calculation of the charmonium spectrum, in the CMRF,
where $P^\mu=(M,\vec 0)$, Eq. (\ref{masscov}) can be written as
\begin{equation}\label{masscovcmrf}
{\cal V}_M^{CMRF}=
 \bar \Psi {\frac {V_M(r)} {2} }
 (\gamma_1^0+\gamma_2^0)  \Psi =
\Psi^\dag {\frac {V_M(r)} {2} }
 (\gamma_1^0+\gamma_2^0)  \Psi ~.
\end{equation}
From the last equation,
we derive the reduced expression of the mass interaction, shown in
in Eqs. (\ref{vmassred})- (\ref{mass2b}) of Appendix \ref{reductm}.
That expression is obtained by using the vinculated
wave functions of our relativistic  model,
applying the reduction operators as shown in Eq. (8) of Ref. \cite{rednumb}.\\
We recall that the reduced expression of the scalar interaction 
was given in Eqs. (C.1)-(C.3) of Ref. \cite{localred}.\\
Considering, for the reduced expressions of the two interactions,
 a ``nonrelativistic"  expansion in powers of
 $p/m$ (being $n$ the power of each term)
one can easily find that the leading term, with $n=0$, is the same for both interactions;
the following terms, with $n=4$, have opposite sign for the two cases.

\vskip 0.5 truecm
\noindent
We recall that for the scalar potential $V_S(r)$  a Gaussian function was used
to fit the charmonium spectrum.
Different functions with the same number of parameters were tested but
the fit to the experimental data strongly favored the Gaussian spatial dependence.
A constant function significantly worsened the reproduction of the data.\\
The author has also tested that
a large distance linear scalar potential (similar to that of the Cornell model)
is unable to reproduce the data with the same accuracy.\\ 
Furthermore,
in work \cite{rednumb} a model was studied with a \textit{two region potential},
with the potential of the outer spatial region of Yukawa form in order to investigate if the scalar interaction can be related to the standard mechanism of 
one scalar meson exchange. 
However, the comparison with 
the experimental data did not show a significant improvement with respect to the Gaussian potential function,
suggesting that the scalar interaction is not originated in this way.
This point will be examined  more deeply also in Sect. \ref{orint}
of this work.
\vskip 0.5 truecm
\noindent

\section{Study of the charmonium spectrum}\label{charmspectr}
In the present work, we study again the scalar interaction
and also consider the mass interaction 
of Eq. (\ref{masscovcmrf}), inserting
the reduced operator $\hat W^M$ of Eq. (\ref{mass2b}) in the RDLE.
On the other hand,
the regularized vector interaction, used in Ref. \cite{rednumb},  is left unchanged. 
We recall that this interaction (that includes the self-energy term) 
is zero at $r=0$ and approaches the value $\bar V_V$ as $r\rightarrow \infty$.
\\
The technique for solving the RDLE
and the fit procedure
are exactly the same as in Ref. \cite{rednumb}.
For the charmonium spectrum we use here the new experimental data \cite{pdg22}
that present some small differences with respect to the old data 
\cite{pdg20} used in  our previous work \cite{rednumb}.
\vskip 0.2 truecm
\noindent
%
For the quality of the fit,
we define
\begin{equation}\label{qual}
\Theta=\sqrt{ {\frac {\sum_k(E_k^{th} -M_k^{exp} )^2} {N_d}} }~,
\end{equation}
where $E_k^{th}$ and $M_k^{exp}$ respectively represent
the result of the theoretical calculation and the experimental value 
of the mass,
for the $k$-th resonance and $N_d=16$ is the number of the fitted resonances.\\
Taking into account the results
obtained in the previous work for the scalar interaction and 
those obtained in the present work, with many different trials,
for both interactions, 
we make the following general comments.
\vskip 0.2 truecm
\noindent
\textit{i}) \textit{Results of the same quality} are  obtained 
with the scalar and mass interaction.
As discussed before, a difference between the scalar and mass reduced interactions
would appear only at the order $n=4$ of a nonrelativistic  expansion. 
In consequence, one can argue that  the motion of the  charm quark and antiquark 
(due to their relatively high mass) is not sufficiently relativistic
to distinguish between  the two interactions. 
We conclude that, at least for the charmonium case, 
both the scalar and the mass interaction are able to reproduce, with high accuracy, the spectrum.\\
\textit{ii}) To obtain a good fit,  $V_X(r)$ must have a Gaussian form,
both for the scalar and for the mass interaction case. 
We can write the x-potential in the form
\begin{equation}\label{vxgauss}
V_X(r)=- \bar V_X \exp \left(- {\frac {r^2} {r_X^2}} \right)~.
\end{equation}
\textit{iii}) A fit of the same quality as that of Ref. \cite{rednumb} is now obtained
 also for the mass interaction, enforcing the same 
\textit{balance equation}, shown in Eq. (23) of Ref \cite{rednumb}.
In this work,
the balance equation will be rewritten, for a general x-interaction, 
in Eqs. (\ref{genbalance1}) and (\ref{genbalance2}) of Sect. \ref{modint}.\\
%
%
\textit{iv}) Finally, the values obtained in the fit for $r_M$, and $\bar V_M$
are numerically very similar to $r_S$ and $\bar V_S$.\\

\vskip 0.2 truecm
\noindent
The results of the spectrum are shown in Table \ref{tabspectr}.
The values of the parameters for the interaction
are given in Table \ref{tabpar}.\\
For the mass of the quark we have taken the same value of the previous work
\cite{rednumb}, that is $m_q=1.27~GeV$.
This value represents the ``running" charm quark mass 
in the $\overline{MS}$ scheme \cite{pdg22}.\\
For reasons that will be explained in Sect. \ref{orint},
when choosing the \textit{free parameters} of the fit, 
we set here a slightly different strategy
with respect to the previous work \cite{rednumb}.
In that work the free parameters were $\alpha_V$, $d$ and $ r_S$,
representing respectively the adimensional coupling constant of the vector
interaction, the regularization distance of the vector interaction and 
the distance parameter of the Gaussian scalar potential.
The dependent parameters were $\bar V_V$ and $\bar V_S$; 
$\bar V_V$ is the two quark vector self-energy that depends on $\alpha_V$
and $d$, as shown in Eq. (16) of that work;
$\bar V_S$ is the two quark scalar self-energy, determined by the balance equation
Eq. (23) of the same work.\\
On the other hand, in the present work we take as \textit{free parameters}
$\bar V_X$, $d$ and $r_X$.
$\bar V_V$ is determined by the balance equation, then $\alpha_V$ is obtained
by means of Eq. (16) of Ref. \cite{rednumb}, as function of $\bar V_V$ and $d$,
that is:
\begin{equation}\label{alphav}
\alpha_V= \sqrt{\pi}   {\frac 3 4}  \bar V_V d~.
\end{equation} 
We now make some comments on the parameter $\bar V_X$.
In our previous work  \cite{rednumb}, the fit procedure with 
the old data \cite{pdg20} 
gave  $\bar V_S=0.7268~GeV$ (see Table II of Ref. \cite{rednumb}).
With the new data \cite{pdg22}, we now obtain $\bar V_S=0.7050~GeV$.
For the case of the mass interaction, we obtain
$\bar V_M=0.7237~GeV$. 
\textit{Instead of these values},
 according to the phenomenological model that will be discussed in Sect. \ref{orint}, 
we give, in Table \ref{tabspectr} and Table \ref{tabpar}, only the results obtained by
fixing $\bar V_X$ (for both $X=S$ and $X=M$) at the value $\bar V_X=0.7350~GeV$,
as it will be discussed in  Sect. \ref{orint};
see, in particular, Eq. (\ref{barvxm0}).
This value is not very different with respect to the results
obtained by  the fit; in consequence,
this choice of $\bar V_X$ (instead of taking the fit results) does not alter significantly 
the reproduction of the  charmonium spectrum. 

\vskip 0.5 truecm
\noindent
Concerning the form of the x-potential,
we consider that the Gaussian form is strongly favored by the fit to the data.
For this reason, in all this work, we focus our attention on the Gaussian
x-potential function:
in the next Section \ref{modint}, we try to develop a microscopic model for this
 interaction
and for  the corresponding contribution to the quark self-energies; 
as anticipated,
in Sect. \ref{orint} we shall discuss a possible phenomenological 
 model  for the origin of this interaction.\\
Finally, we note that the model is unable to reproduce the resonance 
 $\chi_{c0}(3915)$. 
The new experimental data \cite{pdg22} give, for this resonance, a mass of
$3921.7 \pm 1.8 ~MeV$. 
Our model, taking the quantum numbers $2^3P_0$,   
gives  the  mass  values of $3862~MeV$ and $3850~MeV$, for the $S$ 
and the $M$ interactions, respectively.  
Our model and other quark models give a wrong order for the masses of 
this resonance and its partner $\chi_{c1}(3872)$.
On the other hand,
the analysis of the decay processes leaves open the possibility of
different quantum number assignments and of a description
in terms of  multiquark states, as discussed, for example,  in the work \cite{moldes}.
For all these reasons, this resonance has not been included in the fit of
Table \ref{tabspectr}.

\section{A model for the scalar and mass interaction}\label{modint}
In this section we try to develop a  model for the scalar and mass
interaction,  starting from an elementary x-interaction.
This model,
analogously to the vector interaction  case, takes into account a 
finite x-charge distribution of the quarks.
The same form, for the scalar and mass interactions, will be considered\\
We assume that the quarks, 
with an extended distribution of x-charge,
represent the source of the attractive x-interaction. \\
The  procedure that determines the interaction also allows to
analyze in more detail the \textit{self-energy} of those
charge distributions.
This  topic was not examined with sufficient accuracy in our previous
 works where the scalar interaction was only introduced and used to fit 
the charmonium spectrum.\\

\vskip 0.5 truecm
\noindent
As starting point we now make the hypothesis that an \textit{elementary}
x-potential  $V_{EX}(r) $ between two point-like charges
at distance $r$ has the following 
Gaussian form
\begin{equation}\label{expot}
V_{EX}(r)=-\bar V_{EX} \exp \left(- {\frac {r^2} {r_{EX}^2}} \right)~,
\end{equation}
where $r_{EX}$ represents the distance parameter of the elementary x-interaction.
Furthermore, analogously to the vector interaction case,
 we also consider a Gaussian distribution for the quark x-charge density

\begin{equation}\label{rhoxgauss}
\rho_{_X}( x)={\frac {1} {(2 \pi d_X^{^2})^{3/2} }}
\exp\left(-{\frac {\vec x^2} {2{d_X^{^2}} } }\right)~,
\end{equation}
where  $d_X$ is related  the radius  of x-charge distribution density.\\
 In order to construct the global x-potential avoiding cumbersome calculations,
we introduce the corresponding Fourier transformed quantities 
where $\vec q$ represents the vertex momentum transfer and also, $q=|\vec q|$. 
For the potential, transforming Eq. (\ref{expot}),
 we obtain 
 \begin{equation}\label{expotq}
V_{EX}(q)=-\bar V_{EX}
{\frac {r_{EX}^3} {8 \pi ^{3/2}}}
\exp \left(- {\frac {{q^2} {r_{EX}^2}} {4}} \right)~.
\end{equation}
From the x-charge density of Eq. (\ref{rhoxgauss}),
by means of  a Fourier transform, we obtain 
%
the standard vertex form factor 
\begin{equation}\label{formfact}
F_{_X} (q)= 
\exp\left(-~{\frac {{ q^2} {{d_X^{^2}}} } {2} }\right)
\end{equation}
with the normalization $F_X( q=0)=1$.
The global interaction in the $\vec q$ space is obtained
inserting the form factors at the two vertices.
For further developments, we consider the \textit{general} case 
of two different x-charge distributions
for the two sources, with spatial parameters $d_{1 X}$ and $d_{2 X}$.
We have
\begin{equation}\label{vxq}
V_G(q)= F_{_{1X}}( q) V_{EX}(q) F_{_{2X}}( q) = 
-\bar V_{EX}
{\frac {r_{EX}^3} {8 \pi ^{3/2}}}
\exp \left(-  {\frac {q^2  r_G^2} {4} }
\right)~,
\end{equation}
where $r_G$ is defined as:
\begin{equation}\label{rgdef}
r_G=\sqrt{r_{EX}^2 +2( {d_{1X}^2 +d_{2X}^2) }}~.
\end{equation}
Eq. (\ref{vxq}) is easily transformed to the $\vec r$ space. 
The result is
\begin{equation}\label{vxr}
V_G(r )=-\bar V_{EX}\left( {\frac {r_{EX}} {r_G} } \right)^3
\exp \left(- {\frac {r^2} {r_{G}^2}} \right)~.
\end{equation}
Finally, to reproduce the
x-potential of the present model,
 given by Eq. (\ref{vxgauss}),
we take for the two quarks $d_{1X}=d_{2X}=d_{X} $.
By means of Eq. (\ref{rgdef}), we obtain
\begin{equation}\label{rxdef}
r_X=\sqrt{r_{E X}^2 +4 d_{X}^2 }
\end{equation}
and, identifying in Eq. (\ref{vxr})
\begin{equation}\label{identifvx}
\bar V_X=\bar V_{EX}\left( {\frac {r_{EX}} {r_X} } \right)^3~,
\end{equation}
we have 
the same expression for the x-potential  given by
Eq. (\ref{vxgauss}).\\
\vskip 0.5 truecm
\noindent
We now calculate the self-energy of a  spherical  Gaussian 
 x-charge distribution density.\\
To this aim we consider (taking the \textit{general} Eq. (\ref{vxr}))
the potential of a spherical Gaussian distribution
interacting 
with a point-like particle, setting  $d_{1X}=d_X$
and $d_{2X}=0$ in Eq. (\ref{rgdef}).
In this case we have:
\begin{equation}
r_{1 X}=\sqrt{r_{EX}^2 +2 d_{X}^2 }~.
\end{equation}
Defining 
\begin{equation}\label{vbar1x}
\bar V_{1 X}=\bar V_{EX}\left( {\frac {r_{EX}} {r_{1 X}} } \right)^3~,
\end{equation}
we can write this potential as
\begin{equation}\label{v1xr}
V_{1 X}(r )=-\bar V_{1 X}
\exp \left(- {\frac {r^2} {r_{1 X}^2}} \right)~.
\end{equation}
We now replace the point-like x-charge with the x-charge $dQ$ contained in a
 volume element.
For the x-charge density we use the expression of Eq. (\ref{rhoxgauss}). We have:
\begin{equation}
dQ=r^2 dr d \Omega \rho_X(r)~.
\end{equation}
The integration for determining the self-energy is performed  taking into account 
the spherical symmetry of the problem. 
Furthermore, we insert a factor $1/2$ to avoid double counting in the integral
for the total self-energy.
In this way 
the total self-energy of the x-charge distribution of one quark is written in the form
\begin{equation}
W_{1 X}^{self}= {\frac 1 2} 4 \pi {\int_0^\infty} dr  r^2   V_{1 X}(r) \rho_X(r)~.
\end{equation}
A standard calculation finally gives
\begin{equation}
W_{1 X}^{self}={\frac 1 2}V_X(0)= -{\frac 1 2} \bar V_X~.
\end{equation}
Obviously, for charmonium (and $q ~ \bar q$ systems) the \textit{total} self-energy due to the x-interaction is
\begin{equation}\label{wxself}
W_X^{self}= 2 W_{1 X}^{self}= V_X(0)=- \bar V_X~.
\end{equation}

\vskip 0.5 truecm
\noindent
We now comment critically the obtained results.\\
As successfull achievements,
in the first place we note that the Gaussian interaction used in the fit 
can be obtained by means of an \textit{elementary} Gaussian x-interaction and a
a Gaussian x-charge distribution for the quarks.
Moreover, this model determines  the self-energy $W_X^{self}$
 that was used in the balance equation, given in  Eq. (23) of previous work 
\cite{rednumb}.
For clarity we generalize here this equation for a x-interaction:
\begin{equation}\label{genbalance1}
W_V^{self}=2 m_q +W_X^{self}
\end{equation}
or, more explicitly
\begin{equation}\label{genbalance2}
\bar V_V= 2 m_q -\bar V_X~.
\end{equation}

\vskip 0.5 truecm
\noindent
On the other hand, the following two inconveniencies are  found.
\vskip 0.5 truecm
\noindent
\textit{i}) The  negative x-self-energy $W_X^{self}$ of Eq. (\ref{wxself}) 
correctly appears
in the balance equation but gives no contribution to the energies of the charmonium
 spectrum.
Different trials have been performed to include directly this quantity in the potential
functions but the quality of the fit is always greatly worsened.\\
Note that, on the contrary,  for the vector interaction, the self-energy $\bar V_V$
makes part of the regularized potential function and gives $V_V(0)=0$, as shown 
in Eqs. (13)-(17) of Ref. \cite{rednumb}.\\
Some mechanism should be found to cancel the (unwanted) negative x-energy
$W_X^{self}$.
\vskip 0.5 truecm
\noindent
\textit{ii}) As shown in Eq. (\ref{rxdef}) for $r_X$, 
this model as such is unable to determine independently the parameters
$r_{E X}$ and $d_X$, 
introduced respectively in Eqs. (\ref{expot}) and (\ref{rhoxgauss}).
The potential used for the x-interaction only depends on $r_X$, 
that is, in any case, a free fit parameter, determined by 
the charmonium spectroscopy.
From Eq. (\ref{rxdef}) one can only obtain  the following inequalities: 
\begin{equation}\label{ineq}
d_X\leq {\frac {r_X} {2}}~,  ~~~r_E\leq r_X~.
\end{equation}
In the limiting (extreme) case of a point-like x-interaction ($r_E=0$)
we have $d_X= {\frac {r_X} {2}}$.
On the other hand, for a point-like x-charge distribution ($d_X=0$)
we have $r_E=r_X$.\\
\vskip 0.5 truecm
\noindent
To solve these two difficulties and to investigate the physical origin of the Gaussian 
form of the elementary x-interaction, we developed a phenomenological model
that will be discussed in the following section.  

\section{ A model for the origin of the scalar and mass interaction}\label{orint}
As starting point we introduce an underlying Gaussian  field 
for the $q ~\bar q$ system.
This field depends only on 
the interquark distance $r=| \vec r|$;
in this way, it carries a vanishing orbital angular momentum
and can represent a \textit{completely scalar} field with the same 
quantum numbers of the vacuum. 
We shall try to relate this field  to the \textit{elementary} potential $V_{EX}(r)$,
introduced in Eq. (\ref{expot}) of the previous section.
We can write
\begin{equation}\label{phidef}
\Phi(r)=A_{coupl} \cdot \exp \left(- {\frac {r^2} {{\bar r}^2}  }  \right)~,
\end{equation}
where $A_{coupl}$ represents a dimensional coupling constant 
that will be fixed in  the following.\\
In order to understand the dynamical origin of that field,
we apply to $\Phi(r)$ the operator $\vec p ^2= -\vec \nabla^2$, 
obtaining the following equation:
\begin{equation}\label{p2phi}
\left [ \vec p ^2 +  \sigma^2  r^2 \right ]\Phi(r)=
\mu^2  \Phi(r)
\end{equation}
with
\begin{equation}\label{sigma}
\sigma={\frac {2} {\bar r^2}}
\end{equation}
and
  \begin{equation}\label{mu}
	\mu={\frac {\sqrt{6}} {\bar r}}~.
\end{equation}
Examining Eq. (\ref{p2phi}), we note that
the second term in the brackets of the \textit{l.h.s.} 
 represents a 
harmonic   ``potential''
that determines $\Phi(r)$ as a ``confined'' Gaussian field;
in the \textit{r.h.s.} we have the squared energy $\mu^2$
that  can be related to
the mass ($\mu$)  of the \textit{quantum} associated to the Gaussian  field.
We recall that in the case of a vector Coulombic interaction,
the situation is completely different and no term of this kind 
is present.\\
%
At this point, in order to avoid the inconveniency (\textit{i})
found at the end of the previous section,    
we make the hypothesis that
the (positive) value of the quantum $\mu$ 
cancels the negative x-self-energy of the quarks.
To this aim, with the help of Eq. (\ref{wxself}), we fix:
\begin{equation}\label{muwself}
\mu=-W^{self}_X=\bar V_X~.
\end{equation}  
In consequence, from Eq. (\ref{mu}), the parameter $\bar r$ 
of the Gaussian field can be expressed in the form
\begin{equation}\label{barrbarvx}
\bar r={ \frac {\sqrt{6}} {\bar V_X}}~.
\end{equation}

\vskip 0.5 truecm
\noindent
Taking into account the phenomenology of the hadronic interactions,
we note, from Eq. (\ref{muwself}), that the values of $\mu$ obtained 
by fitting the charmonium spectrum 
lie  between the masses of the first two 
scalar  meson resonances that have the vacuum quantum numbers.
More precisely, we recall that for the $f_0(500)$
and the $f_0(980)$ 
the peak of the mass
is roughly estimated at $0.475~GeV$ and at
$0.995~GeV$, respectively \cite{pdg22}.
In consequence, the mean value of these two peaks is, indicatively,
at  $<m_0>=0.7350~GeV$.\\
As anticipated in Sect. \ref{charmspectr},
we  take \textit{this value} for both the cases (X=M or X=S) obtaining
an accurate reproduction of the spectrum.
We have
\begin{equation}\label{barvxm0}
\bar V_X=\mu= <m_0> =0.7350~GeV~.
\end{equation}
\vskip 0.5 truecm
\noindent
From the numerical value of $\mu$ fixed in  the previous equation, 
we also obtain, from Eq. (\ref{barrbarvx}),
$\bar r=0.6575~ fm$ and, from Eq. (\ref{sigma}),
$\sigma= 0.9126     ~ GeV/fm   $.

\vskip 0.5 truecm
\noindent
Phenomenologically, one could consider $\bar V_X=<m_0>$ as an
 input of the model and not as free parameter.
In this way, it is possible to say that the charmonium spectrum
is fitted using \textit{only two} truly free parameters: $d$ and $r_X$,
as shown in Table \ref{tabpar}.
 
\vskip 0.5 truecm
\noindent
We have seen  that our  field $\Phi(r)$ can be phenomenologically related 
 to the mesonic excitations that have the vacuum quantum numbers.
In this sense Eq. (\ref{p2phi}) can be tentatively generalized to give the whole spectrum of these excitations:
\begin{equation}\label{p2phin}
\left [ \vec p ^2 + \sigma^2 r^2 \right ]\Phi_n(r)=
({\cal E}_n)^2  \Phi_n(r)
\end{equation}
 with
\begin{equation}\label{ecaln}
{\cal E}_n={\frac {2} {\bar r}} \sqrt{n+ {\frac 3 2}}~.
\end{equation}
The scalar resonances ($L=0$), that couple to the vacuum,
have $n=0,2,4,....$\\
The case discussed above for Eq. (\ref{p2phi}) corresponds to ground state with $n=0$;
more explicitly, in that case we have $\mu= {\cal E}_0$.\\
We stress that  Eqs. (\ref{p2phin}) and (\ref{ecaln}) are not able, as such,
to describe the spectrum of the scalar resonances.
To this aim one should  take into account further terms in the
``potential" (besides $ \sigma^2 r^2 $) and,
 in any case, consider the quark contributions.
Our model could represent a phenomenological starting point for investigating this highly controversial item of hadronic physics with a different method.
All this subject goes beyond the scope of the present work.
\vskip 0.5 truecm
\noindent
In order to solve the inconvenience (\textit{ii}) of the previous section,
we now study a possible relationship between the Gaussian field $\Phi(r)$ and 
the phenomenological potential $V_X(r)$.\\
We first consider the case of
a \textit{direct coupling} to the fermion fields.
In this case the Gaussian field $\Phi(r)$ of Eq. (\ref{phidef})
represents the elementary x-potential of Eq. (\ref{expot}), that is
\begin{equation}\label{directcoupl}
V_{E X}(r)= \Phi(r)
\end{equation}
that implies
\begin{equation}\label{adirect}
A_{coupl}=A_{dir}= -\bar V_{EX}, ~~~~~~\bar r=r_{_{E X}}~.
\end{equation}
This direct coupling is of the same kind of that studied 
in Refs. \cite{dare1}, \cite{dare2}. 
However, in our model, for the field $\Phi(r)$ no self-interaction term
is considered. 
On the other hand, 
as shown in Eq. (\ref{p2phi}), we have introduced
the ``attractive term" $\sigma^2 r^2$.\\   
By using Eq. (\ref{rxdef}) and $r_{EX}$ from Eq. (\ref{adirect})
it is now possible to determine the distance parameter of
x-charge distribution density:
\begin{equation}\label{detdxdir}
d_X={\frac 1 2}\sqrt{r_X^2-\bar r^2}~.
\end{equation}
Furthermore, by using Eq. (\ref{identifvx}),
the elementary coupling of the x-interaction
can be written as
\begin{equation}\label{detvexdir}
\bar V_{EX}= \bar V_X \left({\frac {r_X} {\bar r}} \right)^3~,
\end{equation}
where, as discussed above, we have $\bar V_X= \mu$.
For completeness, we give the corresponding numerical values.
Using the results of Table \ref{tabpar},
for the scalar interaction, we have:
  $d_S=0.8640~fm$ and $\bar V_{ES}=16.35~GeV$;
for the mass interaction,  we have:
  $d_M=0.8625~fm$ and $\bar V_{EM}=16.27~GeV$.	
\vskip 0.5 truecm
\noindent
Another possibility, at purely phenomenological level,
 consists in taking a \textit{quadratic coupling} of the form
\begin{equation}\label{directcouplq}
V_{E X}(r)= -\Phi^\dag(r) \Phi(r)~.
\end{equation}
In this case
\begin{equation}\label{aquadratic}
A_{coupl}=A_{quad}=\sqrt{\bar V_{EX}}, ~~~~~~\bar r= \sqrt{2}~ r_{_{E X}}~.
\end{equation}
Taking into account the expression of $\bar r$, the distance parameter 
of the x-charge distribution density is obtained from Eq. (\ref{rxdef}), in the form:
\begin{equation}\label{detdxquadr}
d_X={\frac 1 2}\sqrt{r_X^2- {\frac {\bar r^2} {2} } }
\end{equation}
and the elementary coupling of the x-interaction,
from  Eq. (\ref{identifvx}),    can be written as:
\begin{equation}\label{detvexquad}
\bar V_{EX}= \bar V_X \left({\frac {r_{_X} \sqrt{2}} {\bar r}} \right)^3
\end{equation}
with, as before, $\bar V_X= \mu$.
The numerical values, for the scalar interaction, are:
$d_S=0.8948~fm$, $\bar V_{ES}=46.23~GeV$;
for the mass interaction, they are:
$d_M=0.8932~fm$, $\bar V_{EM}=46.01~GeV$.\\
We point out that, in any case, the Gaussian character of the elementary interaction
is essential to obtain the Gaussian spatial x-potential. 
In more detail, we have also tried to use a Yukawa function for the elementary interaction
(as it would be given by the standard linear sigma model)
with Gaussian x-densities for the quarks but the total x-interaction
obtained in this way,
is not able to reproduce, with sufficient accuracy, the charmonium spectrum. 
Moreover, in Ref. \cite{rednumb}, we tried to reproduce the experimental charmonium spectrum
with a ``two region" scalar potential. For the outer region we used 
an exponential function with a spatial decay parameter $r_b=0.7594 ~fm$,
possibly corresponding to a mass $m_b=1/r_b= 0.2598 ~ GeV$.
We observe that this last value cannot be associated to any relevant observable 
hadronic  state.\\
On the other hand, the Gaussian model studied 
in the present work allows for a possible physical interpretation in the
framework of the hadronic phenomenology.
\vskip 0.5 truecm
\noindent
We can now try to summarize the results of the work and to draw some conclusions.
Using a reduced relativistic, energy-dependent, equation, an accurate
reproduction of the charmonium spectrum is obtained with  
a regularized vector interaction and a scalar or mass x-interaction.
For the latter interactions, a Gaussian spatial potential is required to fit the data.
A balance relationship among the quark mass and the vector and x-interaction 
self-energies is established.
In this last section we have shown  that the elementary 
Gaussian x-interaction $V_{EX}(r)$ can be associated to a 
scalar field $\Phi(r)$ 
whose energy quantum $\mu$
cancels the negative self-energy of the x-charge distributions of the quarks.
Phenomenologically, the mass $\mu$ is of the order of the first scalar
hadronic resonances.\\
Furthermore, the scalar field $\Phi(r)$ can be related to $V_{EX}(r)$
by means of a direct or quadratic coupling allowing to determine,
in the two cases,
the distance parameter $d_X$ of the x-charge distribution density of the quarks
and the coupling $\bar V_{EX}$ of the elementary x-interaction. 

\vskip 5.0 truecm


\newpage
\vskip 0.5 truecm
\centerline{{\bf Acknowledgements}}
The author thanks the group of  ``Centro de Excelencia en Computaci\'on Cient\'ifica",
Laboratorio de Biolog\'ia Computacional,
Facultad de Ciencias - Universidad Nacional de Colombia"
for  the computation facilities that were used to perform the numerical calculations of this work.

\appendix
\section{Reduction of the two-body mass interaction}\label{reductm}
For the  mass  interaction 
of Eq. (\ref{masscovcmrf}), we
apply the same procedure used in   Ref. \cite{rednumb}    for the vector and  scalar interactions.
We use the reduction operators $K_1$ and $K_2$,
as in Eq. (8) of Ref. \cite{rednumb}. 
For more generality, we start by taking $U_1(r) \neq U_2(r)$ and obtain

	\begin{equation}\label{vmassred}
\begin{split}
\hat  W^M= 
K_1^\dag K_2^\dag\left[\gamma^0_1 U_1^M(r)+\gamma^0_2 U_2^M(r)\right] K_2 K_1~~~~~~~~~ ~~\\
= U_+^M(r)
               -{\frac {1} {(m_1+E_1)^2 }}
\vec \sigma_1 \cdot \vec p_1      
U_-^M(r)
\vec \sigma_1 \cdot \vec p_1 ~~~~~~~~~~~~\\ 
              + {\frac {1} {(m_2+E_2)^2 }}
\vec \sigma_2 \cdot \vec p_2       
U_-^M(r) 
\vec \sigma_2 \cdot \vec p_2   ~~~~~~~~~~~~\\
%
%
%
%
-{\frac {1} { (m_1+E_1)^2 (m_2+E_2)^2 }}
(\vec \sigma_1 \cdot \vec p_1) (\vec \sigma_2 \cdot \vec p_2)
U_+^M(r)
(\vec \sigma_2 \cdot \vec p_2) (\vec \sigma_1 \cdot \vec p_1)~,
\end{split}
\end{equation}
where we have defined
\begin{equation}
U_\pm^M(r)=U_1^M(r) \pm U_2^M(r) ~.
\end{equation}
For equal mass quarks, we have $m_1=m_2=m_q~,~~ E_1=E_2= E/2$.
Recalling Eq. (\ref{v_m}), one also has
\begin{equation}\label{upmeqmass}
U_-^M(r)=0, ~U_+^M(r)=V_M(r)~.
\end{equation}
Furthermore, 
using $\vec p_2=-\vec p_1=\vec p$
we obtain the following expression that is used for the
calculation of the charmonium spectrum:
\begin{equation}\label{mass2b}
\hat  W^M
=V_M(r)   
-{\frac {1} { (m_q +{\frac E 2})^4 }}
(\vec \sigma_1 \cdot \vec p) (\vec \sigma_2 \cdot \vec p)
V_M(r)
(\vec \sigma_2 \cdot \vec p) (\vec \sigma_1 \cdot \vec p)~.
%
\end{equation}
\newpage


\newpage
\begin{table*}

\caption{Comparison between the experimental average values \cite{pdg22} 
 of the charmonium spectrum (last column)
and the theoretical results of the model.
All the masses are in MeV. 
The quantum numbers  $n$, $L$, $S$ and $J$, introduced in Ref. \cite{rednumb},
respectively
represent the principal quantum number, the orbital angular momentum, the spin 
and the total  angular momentum.
The results of the columns ``Scalar" and ``Mass" respectively refer to 
the scalar (S) and mass (M) interaction.
A line divides the resonances below and above the open Charm threshold.
At the bottom, the quantity $\Theta$, formally in $MeV$, defined in Eq. (\ref{qual}),
 gives an indication of the quality of the fit.  }


\begin{center}
\begin{tabular}{ccccc}
\hline
\hline \\
Name & $n^{2S+1}L_J$  & Scalar &  Mass & Experiment          \\
\hline \\
$\eta_c(1S)$    &  $1^1 S_0 $     & 2998   & 2981    & 2983.9   $\pm$  0.4   \\
$J/\psi(1S)$    &  $1^3 S_1 $     & 3090   & 3102   & 3096.9   $\pm$  0.006 \\
$\chi_{c0}(1P)$ &  $1^3 P_0 $     & 3420  & 3405   & 3414.71  $\pm$ 0.30   \\
$\chi_{c1}(1P)$ &  $1^3 P_1 $     & 3498  & 3497    & 3510.67  $\pm$ 0.05   \\
$ h_c(1P)$      &  $1^1 P_1 $     & 3510  & 3514    & 3525.38  $\pm$ 0.11   \\ 
$\chi_{c2}(1P)$ &  $1^3 P_2 $     & 3564   & 3577    & 3556.17  $\pm$ 0.07   \\
$\eta_c(2S)$    &  $2^1 S_0 $     & 3648   & 3641    & 3637.5   $\pm$ 1.1    \\
$\psi(2S)$      &  $2^3 S_1 $     & 3679  & 3680    & 3686.10 $\pm$ 0.06 \\
\\
\hline \\
$\psi(3770)$&     $1^3 D_1 $     &  3796 & 3795  & 3773.7   $\pm$ 0.4 \\  
$\psi_2(3823)$&   $1^3 D_2 $     &  3831 & 3833  & 3823.7   $\pm$ 0.5  \\
$\chi_{c1}(3872)$&$2^3 P_1 $     &  3893 & 3887  & 3871.65  $\pm$ 0.06 \\
$\chi_{c2}(3930)$&$2^3 P_2 $     &  3928 & 3932  & 3922.5   $\pm$ 1.0  \\
$\psi(4040)$&     $3^3 S_1 $     &  4014 & 4014  & 4039     $\pm$ 1     \\
%
$\chi{c1}(4140)$& $3^3 P_1 $     &  4144 & 4143  & 4146.5  $\pm$ 3.0    \\
$\psi(4230)    $& $4^3 S_1 $     &  4211 & 4216  & 4222.7   $\pm$ 2.6     \\
$\chi{c1}(4274)$& $4^3 P_1 $     &  4267 & 4273  & 4286    $\pm$ 9      \\
\\
\hline
\hline \\
$ \Theta   $          &$ ~$& $13.4 $&$12.8  $& \\  
\\
\hline\\
~\\
~\\
~\\
~\\

\end{tabular}
\end{center}
\label{tabspectr}
\end{table*}

\begin{table*} 
\caption{
Numerical values of the free and dependent parameters of the model; 
$m_q $ is fixed at the value of Ref. \cite{pdg22}, as explained in the text; 
according to the discussion of Sect. \ref{orint},
$\bar V_X$ has the value given by Eq. (\ref{barvxm0}) 
for both scalar and mass case; 
in consequence, $\bar V_V$ has the value  
determined by Eq. (\ref{genbalance2});
$d$ and $r_X$ are the truly free parameters of the model;
finally, $\alpha_V$ is a dependent parameter, determined by Eq. (\ref{alphav}).
         } 
\begin{center}
\begin{tabular}{lllll}
\hline 
\hline \\   
             &             &           &  Units     \\ 
\hline \\
 $m_q $      &  $1.27$     &            &   GeV   \\
\hline \\
$\bar V_X $  &  $0.735$    &            &   GeV   \\
\hline \\
$\bar V_V $  &  $1.805$    &            &   GeV   \\
\hline \\   
             &   Scalar    & Mass       &             \\ 
\hline \\   
 $d        $ &  $0.1511 $  &$0.14045 $  &  fm      \\
 $r_X      $&  $1.849   $  &$1.846   $  &  fm      \\     

\hline\\
$\alpha_V $&  $1.838  $ &$1.708 $ &          \\
\hline

\end{tabular}
\end{center}

\label{tabpar}
\end{table*}

\end{document}